\documentclass[useAMS,usenatbib]{mn2e}

\usepackage{graphicx}
\usepackage{amsmath}
\usepackage{amssymb}
\usepackage{natbib}
\usepackage{aas_macros}
\usepackage{color}
\usepackage{float}
\usepackage{txfonts}

\usepackage{hyperref}
\hypersetup{
    colorlinks,%
    citecolor=blue,%
    linkcolor=blue,%
    urlcolor=blue
}

\title[Polarization in gamma-ray pulsars]{Polarized synchrotron emission from the equatorial current sheet in gamma-ray pulsars}

\author[B. Cerutti et al.]{Beno\^it Cerutti$^{1}$\thanks{E-mail: benoit.cerutti@univ-grenoble-alpes.fr}, J\'er\'emy Mortier$^{1}$ and Alexander A. Philippov$^{2}$\\
$^{1}$Univ. Grenoble Alpes, CNRS, IPAG, F-38000 Grenoble, France\\
$^{2}$Department of Astrophysical Sciences, Princeton University, Princeton, NJ 08544, USA}

\date{Accepted --. Received --; in original form --}

\pubyear{2016}

\begin{document}
\label{firstpage}
\pagerange{\pageref{firstpage}--\pageref{lastpage}}
\maketitle

\begin{abstract}
Polarization is a powerful diagnostic tool to constrain the site of the high-energy pulsed emission and particle acceleration in gamma-ray pulsars. Recent particle-in-cell simulations of pulsar magnetosphere suggest that high-energy emission results from particles accelerated in the equatorial current sheet emitting synchrotron radiation. In this study, we re-examine the simulation data to compute the phase-resolved polarization properties. We find that the emission is mildly polarized and that there is an anticorrelation between the flux and the degree of linear polarization (on-pulse: $\sim 15\%$, off-pulse: $\sim 30\%$). The decrease of polarization during pulses is mainly attributed to the formation of caustics in the current sheet. Each pulse of light is systematically accompanied by a rapid swing of the polarization angle due to the change of the magnetic polarity when the line of sight passes through the current sheet. The optical polarization pattern observed in the Crab can be well-reproduced for a pulsar inclination angle $\sim 60^{\rm o}$ and an observer viewing angle $\sim 130^{\rm o}$. The predicted high-energy polarization is a robust feature of the current sheet emitting scenario which can be tested by future X-ray and gamma-ray polarimetry instruments.
\end{abstract}




\section{Introduction}

Gamma-ray observations show that pulsars are efficient particle accelerators \citep{2010ApJS..187..460A, 2013ApJS..208...17A}. In principle, the exact location of the accelerating regions can be constrained from the spectral and temporal properties of the gamma-ray emission. The detection of high-energy gamma rays $>$GeV in most pulsars pushes the emitting zone away from the polar caps of the star where they would be otherwise absorbed by the magnetic field. The careful analysis of lightcurve morphologies provides another independent constraint which also favors the outer parts of the magnetosphere as the main emitting regions (e.g., \citealt{2010ApJ...715.1270B, 2010ApJ...715.1282B, 2010ApJ...714..810R, 2015A&A...575A...3P}). However, due to our poor knowledge of the pulsar inclination and viewing angles, it turns out to be rather difficult to disentangle between models.

In contrast, the expected polarization signature differs significantly from one model to another \citep{2004ApJ...606.1125D, 2005ApJ...627L..37P, 2007ApJ...670..677T, 2007ApJ...656.1044T, 2013MNRAS.434.2636P} because it is very sensitive to the electromagnetic geometry, and hence to the location of the emitting zones. While the polarization properties in radio (coherent emission) is well-documented (e.g., \citealt{2016JPlPh..82b6301P}), polarization measurements at higher energies (incoherent emission) exist for a few pulsars only (see \citealt{2009MNRAS.397..103S} for a review). The Crab pulsar presents the best multiwavelength coverage, from optical to soft gamma rays \citep{1988MNRAS.233..305S, 2009MNRAS.397..103S, 1996MNRAS.282.1354G, 1978ApJ...220L.117W, 2008Sci...321.1183D, 2008ApJ...688L..29F}. Phase-resolved optical and UV observations report a moderate degree of polarization (PD $\sim 10\%$--$30\%$) with significant swings of the polarization angle for each pulse, which suggests a rapid change in the field geometry.

The equatorial current sheet forming beyond the light cylinder is a natural place for both particle acceleration via magnetic reconnection and sharp changes of the fields because this region separates the two magnetic polarities \citep{1990ApJ...349..538C, 2001ApJ...547..437L, 2005ApJ...627L..37P}. This scenario is supported by global particle-in-cell (PIC) simulations of plasma-filled magnetospheres \citep{2014ApJ...785L..33P, 2015ApJ...801L..19P, 2015ApJ...815L..19P, 2014ApJ...795L..22C, 2015MNRAS.448..606C, 2016MNRAS.457.2401C, 2015MNRAS.449.2759B}. These studies show that $\sim 10$--$20\%$ of the Poynting flux is efficiently dissipated in the current sheet within $1$--$2$ light-cylinder radii and channelled into non-thermal particles acceleration and synchrotron radiation. Pulses of high-energy radiation naturally result from the passage of the current sheet across the observer's line of sight \citep{2016MNRAS.457.2401C}.

In this study, we model the high-energy phase-resolved polarization signal expected in gamma-ray pulsars. In the next section, we present the method to compute the Stokes parameters directly from the PIC simulations. We show a few representative cases as well as a Crab-like configuration in Section~\ref{sect_results}. We briefly discuss our results in Section~\ref{sect_discussion}.

\section{Methods}\label{sect_methods}

This study is based on the 3D global PIC simulations by \citet{2016MNRAS.457.2401C} performed with the {\sc zeltron} code \citep{2013ApJ...770..147C}, from which we compute the polarization properties. The pulsar is modelled as a rotating dipole whose magnetic moment ($\boldsymbol{\mu}$) is inclined at an angle $\chi$ with respect to the star angular velocity vector ($\boldsymbol{\Omega}$). The simulation box extends from the neutron star surface $r_{\rm min}=r_{\star}$ up to $r_{\rm max}=3 R_{\rm LC}$, where $R_{\rm LC}=c/\Omega=3r_{\star}$ is the light-cylinder radius. Self-consistent pair production is not considered in this simulation. Instead, the magnetosphere is fed with low-energy electron-positron pairs continuously injected at the surface of the star. The plasma density is high compared with the local Goldreich-Julian density \citep{1969ApJ...157..869G}. Thus, the magnetosphere is almost force-free everywhere, except in the equatorial current sheet where magnetic reconnection accelerates particles. In addition to the Lorentz force, the particles are subjected to the radiation reaction force to account for the strong curvature and synchrotron cooling. Here, we focus on the high-energy particles only (i.e., with a Lorentz factor $\gamma>10$) which are responsible for the synchrotron radiation emitted within the equatorial current sheet. For more details about the simulations, see \citet{2016MNRAS.457.2401C}. 

\begin{figure}
\centering
\includegraphics[width=8.5cm]{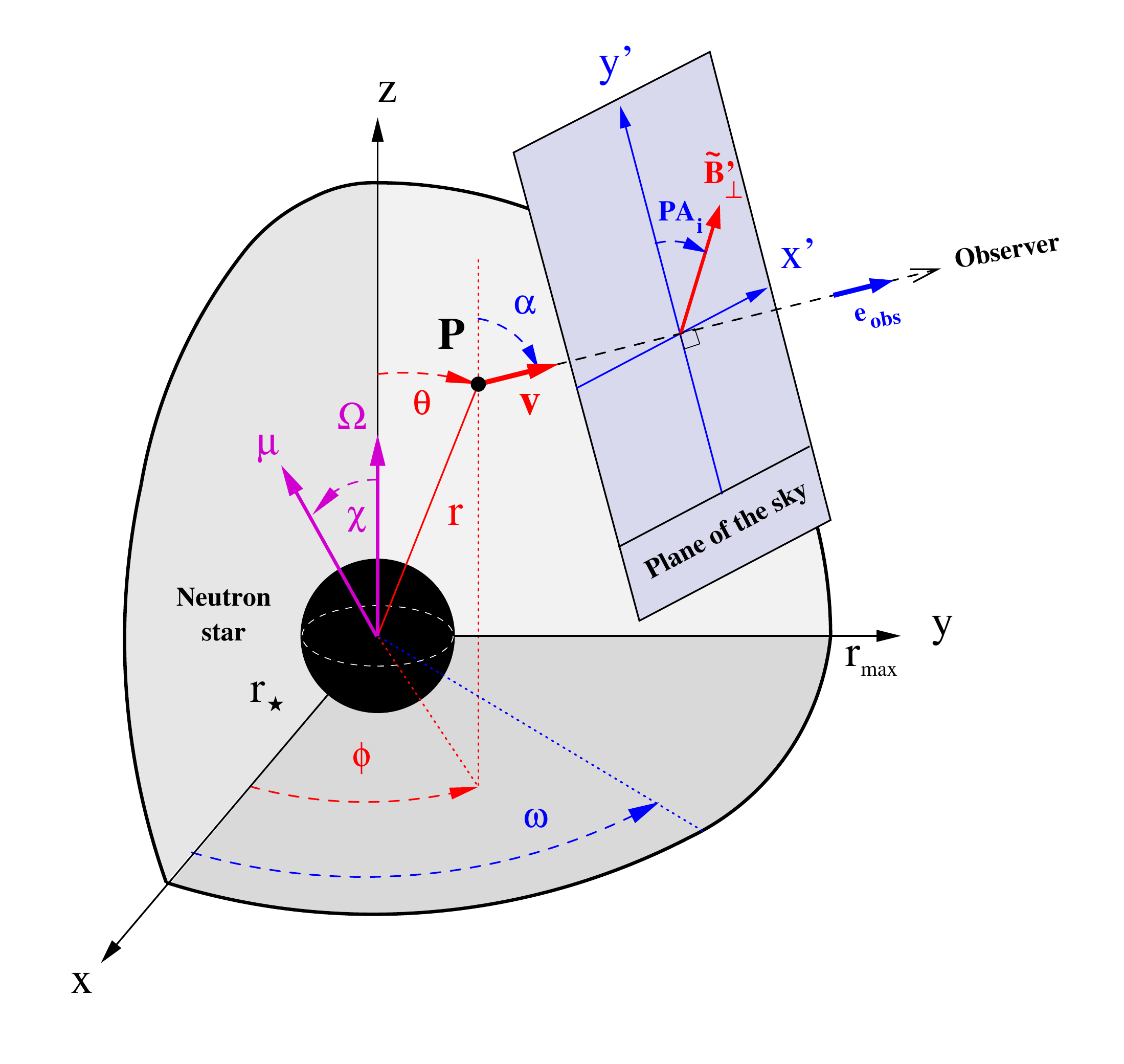}
\caption{Geometry of the problem. A particle located in P$(r,\theta,\phi)$ emits synchrotron radiation boosted along its direction of motion ($\mathbf{v}$) and along the line of sight ($\mathbf{e_{\rm obs}}$). The angle between the $y'$-axis and the vector $\mathbf{\tilde{B}_{\perp}}=\mathbf{E} + \boldsymbol{\beta}\times\mathbf{B} -\left(\boldsymbol{\beta} \cdot \mathbf{E}\right) \boldsymbol{\beta}$ measured in P and projected on the plane of the sky ($\mathbf{\tilde{B}^{\prime}_{\perp}}$) is the angle of polarization PA$_{i}$ of the light emitted by the particle $i$.}
\label{fig_geo}
\end{figure}

The modelling of polarization is done as follows. Each simulation particle emits photons propagating along the particle direction of motion. The angular spread of the emission is neglected which is a good approximation for ultrarelativistic particles. Once emitted, the photons are not reabsorbed unless they are heading towards the star in which case they are removed from the simulation. Photons are then collected on a screen located at infinity as a function of the viewing angle ($\alpha$), the normalized pulsar phase\footnote{The plane containing $\boldsymbol{\mu}$ and $\boldsymbol{\Omega}$ defines the origin of phases.} ($\Phi_{\rm P}$), and the frequency ($\nu$) taking into account the time of flight between the emitter and the observer, $t_{\rm d}$. For a particle located in $(r,\theta,\phi)$ and emitting towards an observer located in the direction ${\bf e_{\rm obs}}$, then $t_{\rm d}=-\left(\mathbf{r}\cdot\mathbf{e_{\rm obs}}\right)/c$. The plane of the sky is perpendicular to the direction of the observer. In this plane, we define two perpendicular directions $x'$ and $y'$, in such a way that $y'$ is aligned with the projection of the pulsar rotation axis on the sky. In the fixed Cartesian coordinates system $(x,y,z)$ shown in Figure~\ref{fig_geo}, the unit vectors along $x'$ and $y'$ and the observer are
\begin{equation}
\mathbf{\hat{x}'}=\begin{pmatrix}
         -\sin\omega \\
         \cos\omega \\
         0
\end{pmatrix}
,\hspace{0.1cm}
\mathbf{\hat{y}'}=\begin{pmatrix}
         -\cos\alpha\cos\omega \\
         -\cos\alpha\sin\omega \\
         \sin\alpha
\end{pmatrix}
,\hspace{0.1cm}
\mathbf{e_{\rm obs}}=\begin{pmatrix}
         \sin\alpha\cos\omega \\
         \sin\alpha\sin\omega \\
         \cos\alpha
\end{pmatrix}
.
\end{equation}
To compute polarization, we measure the orientation of the vector (see \S17 in \citealt{1975ctf..book.....L})
\begin{equation}
\mathbf{\tilde{B}_{\perp}}=\mathbf{E} + \boldsymbol{\beta}\times\mathbf{B} -\left(\boldsymbol{\beta} \cdot \mathbf{E}\right) \boldsymbol{\beta}
\label{eq_Bperp}
\end{equation}
at the particle position and projected on the plane of the sky, where  $\mathbf{E}$ and $\mathbf{B}$ are the electric and magnetic fields, and $\boldsymbol{\beta}=\mathbf{v}/c$ is the particle 3-velocity divided by the speed of light\footnote{If $\mathbf{E}=\mathbf{0}$, one recovers the usual synchrotron case where only $\boldsymbol{\beta}\times\mathbf{B}$ is projected on the sky.}. Then, the polarization angle of the light emitted by the particle $i$ is given by
\begin{equation}
\cos{\rm PA}_i=\frac{\mathbf{\tilde{B}_{\perp}}\cdot \mathbf{\hat{y}'}}{\sqrt{\left(\mathbf{\tilde{B}_{\perp}}\cdot\mathbf{\hat{x}'}\right)^2 + \left(\mathbf{\tilde{B}_{\perp}}\cdot\mathbf{\hat{y}'}\right)^2}}
\end{equation}
\begin{equation}
\sin{\rm PA}_i=\frac{\mathbf{\tilde{B}_{\perp}}\cdot \mathbf{\hat{x}'}}{\sqrt{\left(\mathbf{\tilde{B}_{\perp}}\cdot\mathbf{\hat{x}'}\right)^2 + \left(\mathbf{\tilde{B}_{\perp}}\cdot\mathbf{\hat{y}'}\right)^2}},
\end{equation}
where ${\rm PA}_i$ is defined with respect to the $y'$-axis in the interval $\left[0,\pi\right]$ (Figure~\ref{fig_geo}).

The Stokes parameters can then be reconstructed by summing over the contributions from all the particles pointing towards the observer, $N_{\rm obs}$, at a given pulsar phase \citep{1965ARA&A...3..297G}
\begin{equation}
I = \sum_{i=1}^{N_{\rm obs}} w_i F\left(\xi_i\right),\hspace{1.25cm}
\end{equation}
\begin{equation}
Q = \sum_{i=1}^{N_{\rm obs}} w_i G\left(\xi_i\right)\cos \left(2\rm{PA}_{i}\right),
\end{equation}
\begin{equation}
U = \sum_{i=1}^{N_{\rm obs}} w_i G\left(\xi_i\right)\sin \left(2\rm{PA}_{i}\right),
\end{equation}
where $w_i$ is the particle weight (i.e., the number of physical particles each PIC particle represents), $\cos \left(2\rm{PA}_{i}\right)=\cos^2{\rm PA}_i-\sin^2{\rm PA}_i$,  $\sin \left(2\rm{PA}_{i}\right)=2\sin{\rm PA}_i\cos{\rm PA}_i$, $F(\xi)=\xi\int_\xi^{+\infty}K_{5/3}(\xi')d\xi'$ and $G(\xi)=\xi K_{2/3}(\xi)$ are the usual functions associated with synchrotron radiation. The parameter $\xi_i=\nu/\nu_{\rm c}$ is the radiation frequency divided by the critical synchrotron frequency defined as
\begin{equation}
\nu_{\rm c}=\frac{3 e \lVert\mathbf{\tilde{B}_{\perp}}\rVert \gamma^2}{4\pi m_{\rm e}c},
\end{equation}
where $\gamma=1/\sqrt{1-\beta^2}$ is the particle Lorentz factor, $e$ is the elementary electric charge and $m_{\rm e}$ is the electron mass. The last Stokes parameter is negligible for ultrarelativistic particles, $V\approx 0$, i.e., the radiation is linearly polarized. The first Stokes parameter ($I$) is the total intensity (polarized and unpolarized) of the light. The degree of linear polarization is given by
\begin{equation}
{\rm PD}=\frac{\sqrt{Q^2+U^2}}{I}.
\end{equation}
The total polarization angle (defined with respect to the $y'$-axis between 0 and $\pi$) is obtained from
\begin{equation}
\tan \left(2\rm{PA}\right)=\frac{U}{Q},
\end{equation}
so that
\begin{equation}
{\rm PA} = \frac{1}{2}\rm{atan2}\left(U,Q\right),~{\rm if}~U>0
\end{equation}
\begin{equation}
{\rm PA} = \pi+\frac{1}{2}\rm{atan2}\left(U,Q\right),~{\rm if}~U<0.
\end{equation}

\begin{figure*}
\centering
\includegraphics[width=8.75cm]{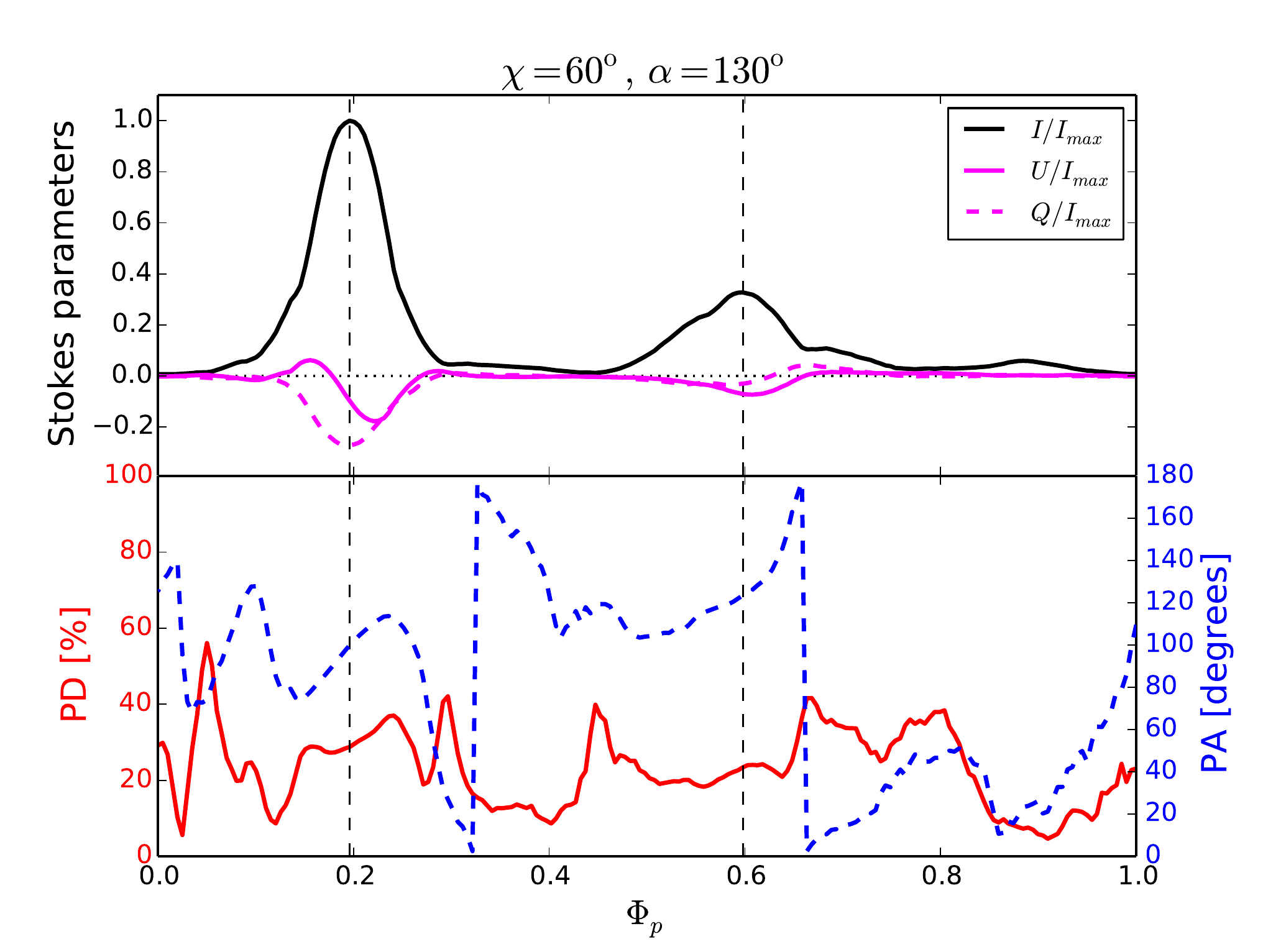}
\includegraphics[width=8.75cm]{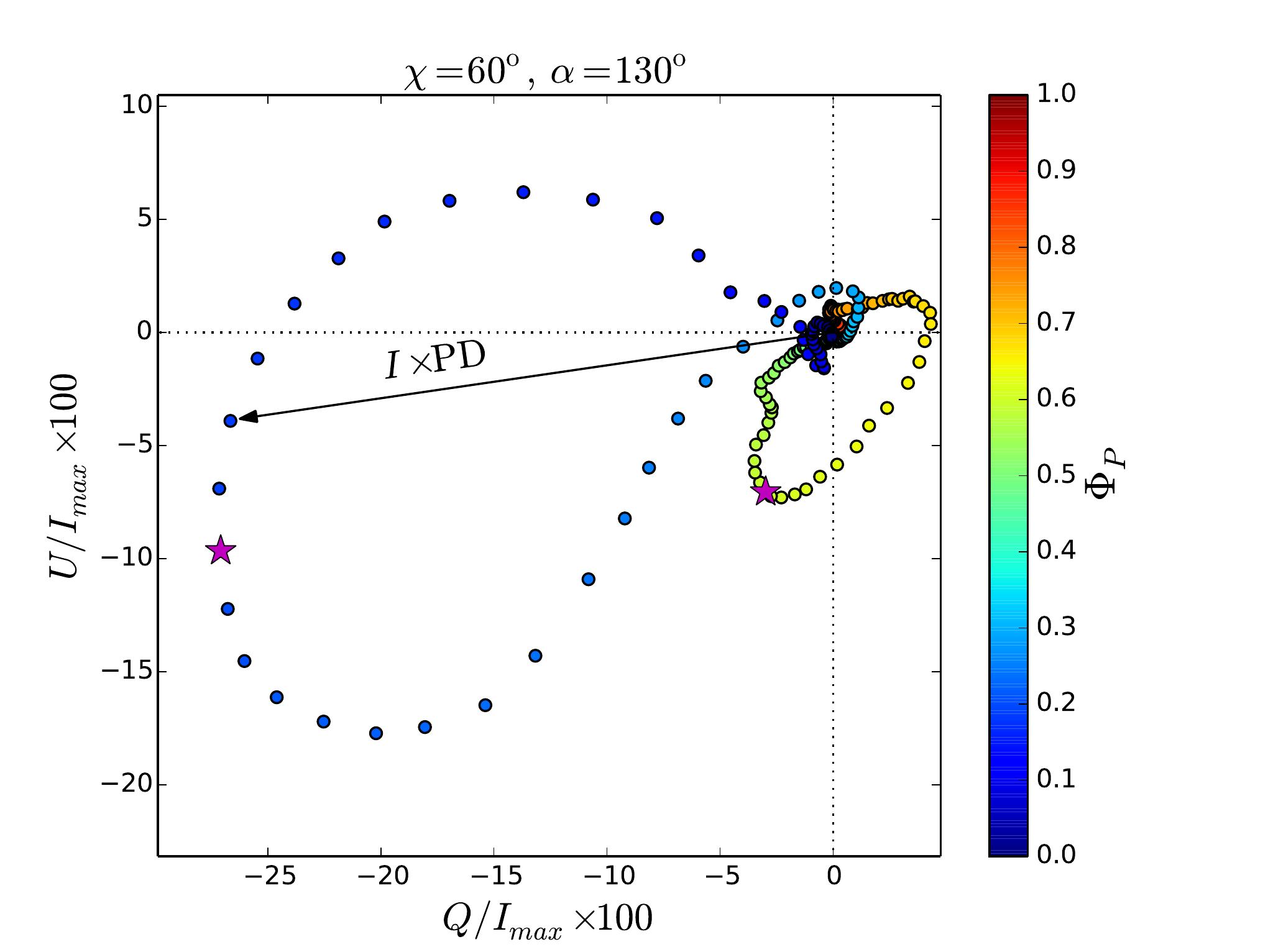}
\includegraphics[width=8.75cm]{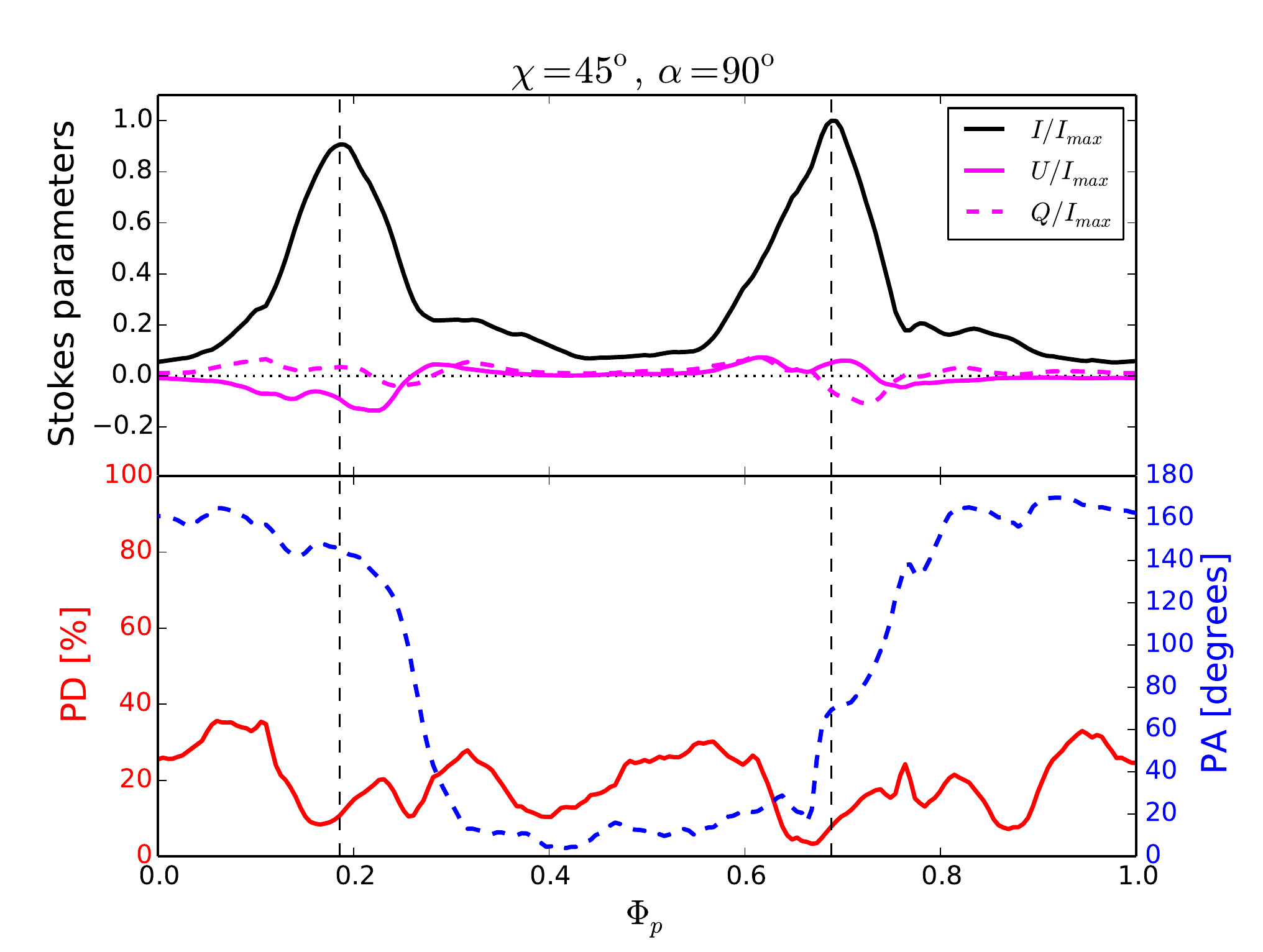}
\includegraphics[width=8.75cm]{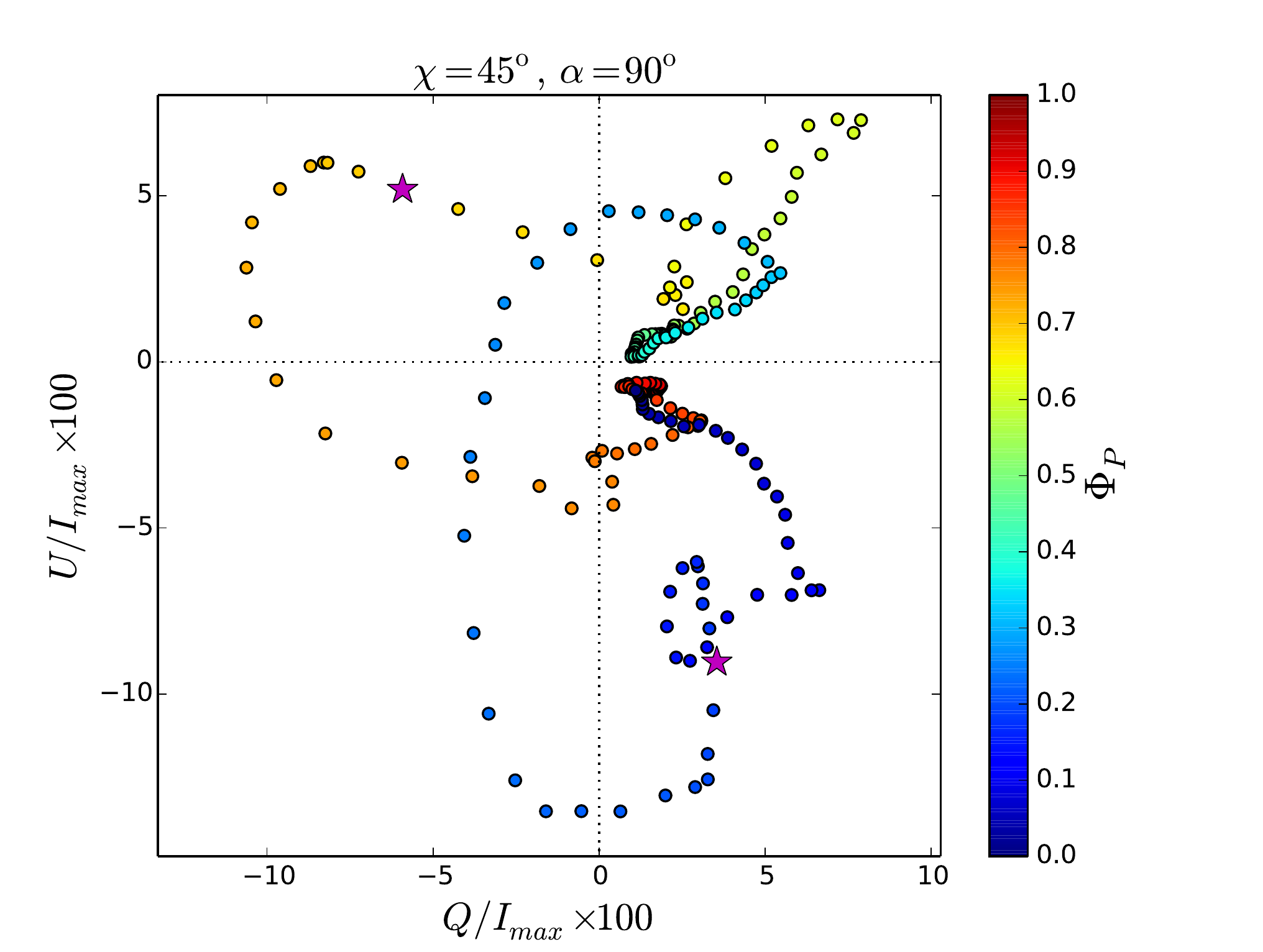}
\includegraphics[width=8.75cm]{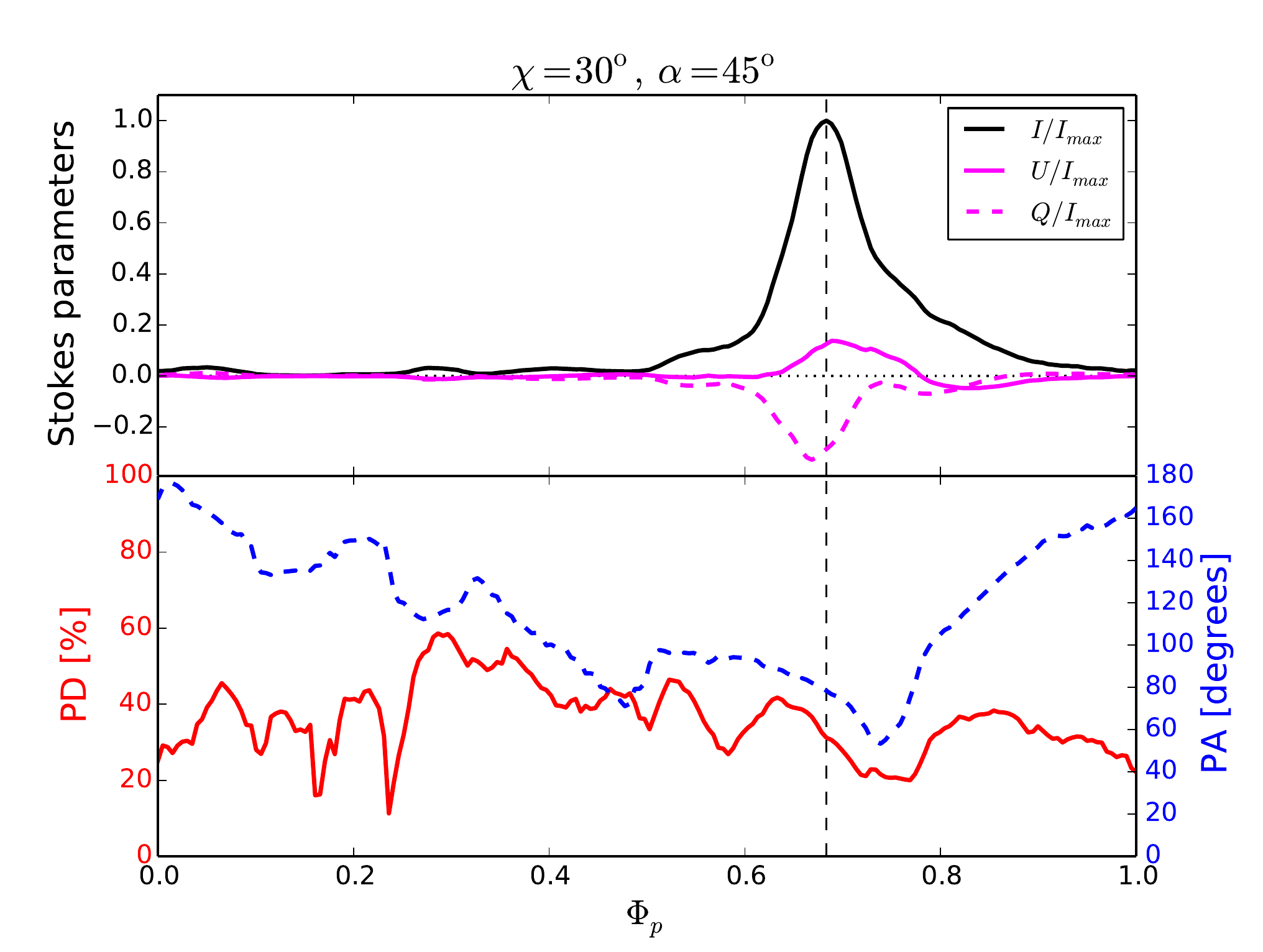}
\includegraphics[width=8.75cm]{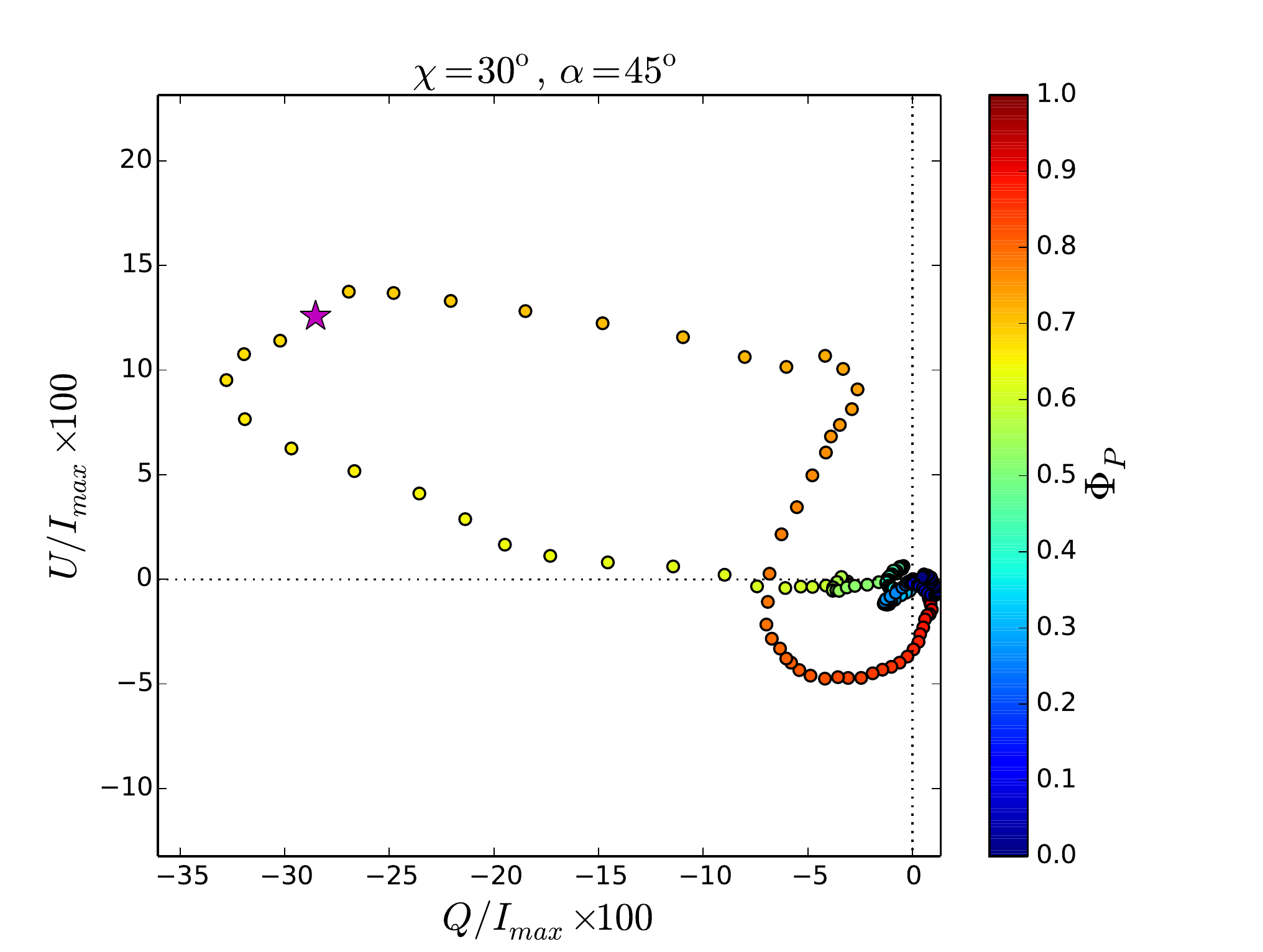}
\caption{Calculated polarization properties for a few representative solutions: $\chi=60^{\rm o}$ $\alpha=130^{\rm o}$ (Crab-like lightcurve, top panels), $\chi=45^{\rm o}$ $\alpha=90^{\rm o}$ (symmetric double-peaked lightcurve, middle panels), $\chi=30^{\rm o}$ $\alpha=45^{\rm o}$ (single-peak lightcurve, bottom panels). Left panels show the Stokes parameters ($I$, $Q$, $U$) normalized to the lightcurve maximum $I_{\rm max}$, the degree of polarization (PD in percent) and the angle of polarization (PA in degrees). Right panels: corresponding vector diagram $U(Q)$. The color indicates the pulsar phase $\Phi_{\rm P}$. The maximum of each pulse of light is shown by a magenta star. Note that the PD and PA curves are sensitive to a background polarized component while the $U(Q)$ diagram would just be offset from the origin. Note that the Stokes parameters were smoothed with a Savitzky-Golay filter.}
\label{fig_pol}
\end{figure*}

\section{Results}\label{sect_results}

We generate synthetic lightcurves and phase-resolved polarization properties as function of the pulsar viewing angle and pulsar obliquity. We smoothed the Stokes parameters using a Savitzky-Golay filter to remove the short wavelength noise inherent to PIC simulations (see Figure 10 in \citealt{2016MNRAS.457.2401C} for the raw lightcurves). Instead of an exhaustive presentation of the whole parameter space, we have selected a few representative examples shown in Figure~\ref{fig_pol} that we discuss in details below.

The first case presented here is a Crab-like solution (top panels). The observed lightcurve has two peaks separated by about $\Delta \Phi_{\rm P}\approx 0.4$ in phase with some bridge emission in between \citep{2010ApJ...708.1254A}. The first peak is about three times brighter than the second peak. The best solution to match all these properties is for an inclination $\chi=60^{\rm o}$ and a viewing angle $\alpha\approx 130^{\rm o}$. This solution is shown in the top panel of Figure~\ref{fig_pol}, along with the Stokes parameters $Q$ and $U$. The degree of polarization is moderate, about $\rm{PD}\approx 20\%$ on average. It presents significant variations with the pulsar phase of $\pm 10\%$ but without obvious correlations with the pulses. The angle of polarization PA swings significantly at each pulse, i.e., after each passage across the current layer when the magnetic polarity flips. We note that the phase-resolved PD and PA curves shown here are very sensitive to any background polarized component (e.g., coming from the nebula), while the variations of $Q$ and $U$ are unaffected (it adds an offset).

The rapid changes of the polarization angle can be best visualized in the vector diagram (Figure~\ref{fig_pol}, right panel) where each loop corresponds to a swing. In this plot, each dot represents a point at the pulsar phase given by the colorbar. The distance from the origin is $I\times$PD, and the angle to the $Q$-axis is 2PA. Both loops are pointing in the same quadrant ($U<0$ and $Q<0$ for both pulse maximum), which is a key feature of this solution. Although not identical in details, the calculated morphology of the vector diagram is similar to the observed ones in optical and UV \citep{1988MNRAS.233..305S, 1996MNRAS.282.1354G, 2009MNRAS.397..103S}. The main difference is that the two loops are slightly offset from each other while observations show that the small loop (second peak) is entirely contained within the big loop (first peak).

The second case corresponds to a lightcurve with two bright symmetric peaks separated by $\Delta\Phi_{\rm P}\approx 0.5$ in phase for an inclination $\chi=45^{\rm o}$ and viewing angle $\alpha=90^{\rm o}$ (Figure~\ref{fig_pol}, middle panels). This solution is also characterized by large swings of the polarization angle correlated with the pulses. In contrast to the Crab-like solution, the loops in the vector diagram are oriented in opposite directions. This configuration presents a mild correlation between the pulses and a dip in the degree of polarization (on-pulse: PD$\sim 5$--$10\%$, off-pulse: PD$\sim 20$--$30\%$). The last case shown here is a lightcurve with a single pulse at phase $\Phi_{\rm P}\sim 0.7$ for $\chi=30^{\rm o}$ and $\alpha=45^{\rm o}$ (Figure~\ref{fig_pol}, bottom panels). Here again, there is a swing of PA associated with the pulse of light, although we note a smooth change spread over all phases. The degree of polarization is about $40\%$ off-pulse with a dip at $20\%$ close to the peak at phase $\Phi_{\rm P}\approx 0.75$. 

\begin{figure}
\centering
\includegraphics[width=8.75cm]{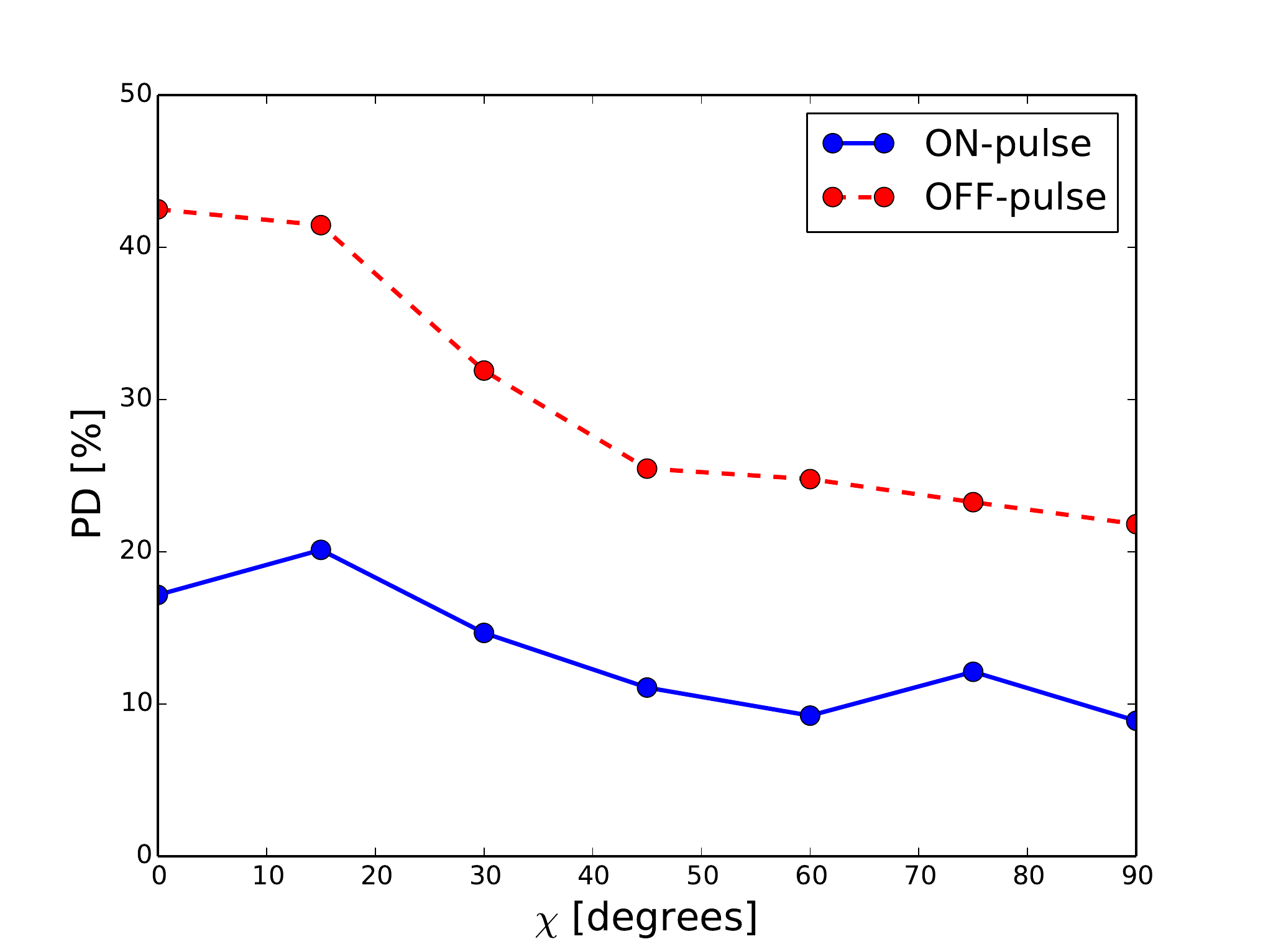}
\caption{On-pulse and off-pulse degree of polarization averaged over the pulsar phase and viewing angle as a function of the pulsar obliquity. On-pulse regions are defined where $I>0.1I_{\rm max}$ for each lightcurve.}
\label{fig_onoff}
\end{figure}

The correlation between high flux and low degree of polarization is clearly seen in Figure~\ref{fig_onoff}. This figure shows the degree of polarization averaged over the pulsar phase and viewing angle as a function of the pulsar inclination. We have arbitrary defined the on-pulse regions where the $I>0.1I_{\rm max}$ (where $I_{\rm max}$ is the maximum of each lightcurve), the rest being the off-pulse regions. On-pulse regions are on average $10$--$20\%$ polarized while the off-pulse is $20$--$40\%$ polarized, depending on the pulsar inclination.

\section{Discussion and conclusion}\label{sect_discussion}

We report on the self-consistent modelling of phase-resolved polarization of the incoherent pulsed emission in gamma-ray pulsar. The expected synchrotron radiation emitted by the equatorial current sheet is mildly polarized at a $\sim 15\%$ level on-pulse and $\sim 30\%$ off-pulse depending on the pulsar viewing angle and magnetic inclination. In most cases, there is a clear anticorrelation between the total observed flux and the degree of polarization, as also noted previously by \citet{2004ApJ...606.1125D} but in the context of the two-pole caustic model. Although the emitting regions are different in their model, the origin of the depolarization is similar here: it is due to the formation of caustics in the observed emission pattern. The lightcurve peaks are formed of photons emitted in different parts of the current sheet, and hence with different magnetic geometries, arriving in phase towards the observer \citep{2016MNRAS.457.2401C}. The resulting signal is depolarized by the superposition of the different components. This effect is particularly severe here because most of the emission occurs within $1$--$2 R_{\rm LC}$ where the orientation of the field changes rapidly (from a poloidal- to toroidal-dominated structure). The small scale turbulence in the current sheet generated by kinetic instabilities (tearing and kink modes, \citealt{2015ApJ...801L..19P, 2016MNRAS.457.2401C}) may also contribute to depolarize the emission. The other robust feature emerging from this study is the sudden swing of the polarization angle (by almost $180^{\rm o}$, i.e., visible as a loop in the vector diagram) coincident with each pulse of light. The swings can be interpreted as the change of magnetic polarity when the observer's line of sight crosses the current sheet \citep{2005ApJ...627L..37P}.

The calculated polarization signatures are qualitatively in agreement with the phase-resolved optical data of the Crab pulsar \citep{2009MNRAS.397..103S}, for a pulsar inclination angle $\chi\sim 60^{\rm o}$ and viewing angle $\alpha\sim 130^{\rm o}$ consistent with the usual estimates inferred from the X-ray morphology of the nebula (e.g., \citealt{2012ApJ...746...41W}). This is the only solution which presents both the correct lightcurve morphology and the correct polarization pattern. However, there are differences in the phase-resolved PD and PA curves between the model and observations that we attribute to a background polarized component that is difficult to subtract \citep{2009MNRAS.397..103S}. We can recover a qualitative agreement by adding a constant polarized emission at a few percent level. The striking resemblance between the optical and gamma-ray lightcurves as well as the spectral continuity between these two bands in the Crab suggest a similar origin of the incoherent radiation. The optical emission may be radiated by low-energy pairs produced in the current sheet \citep{1996A&A...311..172L}. In addition, the close similarity between the optical and UV polarization data from the Crab implies that the polarization properties is not strongly frequency-dependent \citep{1996MNRAS.282.1354G}. Hence, one might expect a similar polarized emission at even higher energies as suggested by our results. If our predictions are correct, $10$--$20\%$ of gamma-ray polarization with the {\em Fermi}-LAT may be detectable for the brightest pulsars like Vela (R. Buehler, private communication). Future X-ray and gamma-ray missions dedicated to polarimetry will provide valuable constraints and tests of the current sheet emitting scenario.

\section*{Acknowledgements}

We thank G. Dubus, G. Henri, A. Spitkovsky, and the referee B. Rudak for valuable comments. This work was supported by CNES and Labex OSUG@2020 (ANR10 LABX56). This research was supported by the NASA Earth and Space Science Fellowship Program (grant NNX15AT50H to AP). The authors acknowledge the Texas Advanced Computing Center (TACC) at The University of Texas at Austin for providing HPC resources via the XSEDE allocation TG-PHY140041 that have contributed to the results reported within this Letter.




\bibliographystyle{aa}
\bibliography{pulsar_polarization} 


\bsp	
\label{lastpage}
\end{document}